# Hyperbolic Metamaterial Feasible for Fabrication with Direct Laser Writing Processes

Xu Zhang[1], Sanjoy Debnath[1], and Durdu Ö. Güney[1,*]

[1] *Michigan Technological University, Department of Electrical and Computer Engineering, 1400 Townsend Drive, 49931, Houghton, Michigan, USA*
*Corresponding author: [*]dguney@mtu.edu

Stimulated emission depletion microscopy inspired direct laser writing (STED-DLW) processes can offer diffraction-unlimited fabrication of 3D-structures, not possible with traditional electron-beam or optical lithography. We propose a hyperbolic metamaterial for fabrication with STED-DLW. First, we design meandering wire structures with three different magnetic dipoles which can be excited under different incidences of light. Then, based on effective parameters corresponding to normal incidence and lateral incidence, we find that the hyperbolic dispersion relation for five-layer structure appears between 15THz to 20 THz. Finally, we investigate the influence of imaginary parts of the effective parameters on the metamaterial dispersion. The proposed metamaterial structure has also potential for three-dimensionally isotropic permeability despite geometric anisotropy. © 2015 Optical Society of America

http://dx.doi.org/10.1364/AO.99.099999

## 1. Introduction

The field of electromagnetic metamaterials has provided us with a new look at the materials by mimicking nature through electromagnetic engineering in subwavelength scales. This has led to the possibility of previously unthought-of applications such as flat lens [1], perfect lens [2], hyperlens [3-6], ultimate illusion optics [7-9], perfect absorber [10, 11], quantum levitation [12], optical analogue simulators [13-15], compact antennas [16, 17], solar photovoltaics [18], metaspacers [19], and many others.

Hyperbolic metamaterials [20-22] have emerged as one of the most interesting and promising subclasses of metamaterials after negative index metamaterials [1,23-25] with practical applications ranging from subwavelength imaging [3-6] to the engineering of spontaneous [26-34] and thermal emission [35].

In 2006, a far-field optical lens with a resolution beyond the diffraction limit was theoretically proposed [3]. This lens was called "hyperlens," because the key feature of the lens enabling subwavelength resolution arises from the hyperbolic dispersion of the metamaterials from which the lens is built. The hyperlens allows image magnification and is more robust with respect to losses as opposed to Pendry's perfect lens [2]. In 2007, the hyperlens was experimentally demonstrated [4]. The fabricated structure showed a subdiffraction resolution of 130nm under 365nm ultraviolet (UV) illumination. In 2010, a spherical version of the hyperlens operating at 410nm visible wavelength and resolving features down to 160nm was designed and fabricated [6]. This was the first experimental demonstration of a far-field imaging device at a visible wavelength, with resolution beyond the diffraction limit in two lateral dimensions.

Furthermore, it was predicted theoretically and demonstrated experimentally that spontaneous emission rate of a dipole emitter would be significantly enhanced near or inside the material with hyperbolic dispersion due to high photonic density of states [26-29, 34]. In an experiment a multilayer metamaterial with hyperbolic dispersion was used for the demonstration of broadband enhanced spontaneous emission from nitrogen-vacancy centers in nanodiamonds [31]. Hyperbolic metamaterials incorporating quantum emitters were proposed for efficient and directional single photon sources for potential applications in quantum information [30]. Roughened surfaces of hyperbolic metamaterials consisting of silver nanowire arrays grown in alumina membranes were demonstrated to scatter light preferentially inside the metamaterial [32]. In a similar experiment, multi-layer hyperbolic metamaterial covered with indium-tin-oxide nanoparticles was shown to reduce reflection [33], which may be useful for high-efficiency solar cells and photodetectors. Besides spontaneous emission engineering by controlling the photonic density of states, hyperbolic dispersion was also shown to provide broadband thermal emission beyond the black-body limit in the near field due to the thermal excitation of unique bulk metamaterial modes [35]. Other found implications of hyperbolic dispersion are extremely high field enhancement [30] and giant optical forces [31] in waveguide.

Among the natural materials, triglycine sulfate and sapphire exhibit hyperbolic dispersion at far-infrared frequencies, bismuth at THz frequencies, and graphite at UV frequencies [38, 39]. To date, most optical metamaterial structures have been fabricated by well-established two-dimensional (2D) fabrication technologies, such as e-beam lithography and evaporation of metal films. However, these can only allow stacking of several planar functional layers [40, 41]. Concerning the fabrication of optical hyperbolic metamaterials, layered metal-dielectric structures [4, 6, 28, 29, 31, 33, 42-45] and nanowire arrays [26, 32, 46-50] have appeared as two common approaches. The largest sample size of 1cm×1cm×51μm was achieved with nanowire arrays fabricated by electrochemical deposition of a metal on a porous alumina membrane [41]. Multilayer fishnet structures [51] and graphene metamaterials [52, 53] with hyperbolic dispersion have been theoretically proposed. However, to fabricate truly bulk optical metamaterials [54, 55] a 3D fabrication approach is needed. Particularly, the practical realization of hyperbolic metamaterial devices such as hyperlens, which is one of the most captivating manifestation of hyperbolic dispersion, demands three-dimensional volume structures.

Direct laser writing (DLW), based on two-photon polymerization, can enable the fabrication of truly bulk and computer controlled arbitrarily shaped three-dimensional (3D) complex structures [56-59] that are not possible with traditional photolithographic processes [60]. DLW has an important potential in fabrication of metamaterials, especially at frequencies ranging from mid-IR to visible, since it offers a viable route as a low-cost and rapid prototyping tool for truly 3D fabrication of nanostructures. Fabrications of large-area, complex metallic nanostructures [61] and metamaterials [57, 58] have been demonstrated with DLW and subsequent metallization. With stimulated-emission-depletion-microscopy inspired direct laser writing (STED-DLW) [62-65], a feature size reduction by more than a factor of two has been demonstrated [66]. Additionally, with the combination of STED-DLW and the "dip-in" approach [67, 68] metamaterial height can reach the level of 1mm—where one can think about constructing macroscopic metamaterials [68].

Here, inspired by the rapid progress in DLW technologies, we propose the first blueprint of a hyperbolic metamaterial structure amenable to fabrication with STED-DLW processes followed by electroplating of gold [58]. The structure has operating frequencies at mid-IR frequencies and the features within the resolution of state-of-the-art STED-DLW technologies.

## 2. Physical geometry

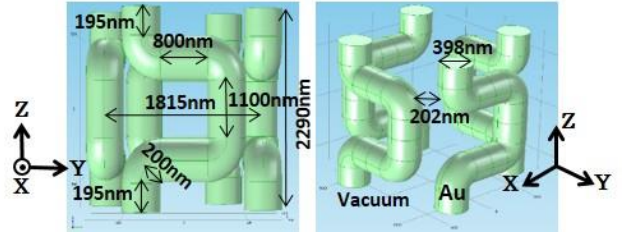

Fig. 1. Free standing gold hyperbolic metamaterial unit cell structure with geometric parameters.

The unit cell of the hyperbolic metamaterial structure consists of two pairs of meandering wires with inversion symmetry (see Fig. 1). The dimension of the unit cell is 2513nm × 2513nm × 2290nm. The wires are modeled by using experimental Drude model parameters for bulk gold with plasma frequency of $f_p$=2180THz and collision frequency of $f_c$=19.1THz as given in [58]. The simulations are performed by using finite integration method based CST Microwave Studio software package. Frequency domain solver is used to calculate the s-parameters corresponding to the complex reflection and transmission coefficients. Then, these s-parameters are used to retrieve the effective medium parameters of the metamaterial [69]. Unit cell boundary conditions are chosen to impose the periodic or quasi-periodic boundary conditions as necessary in the simulations. The tetrahedral meshes with adaptive meshing method are selected to accurately represent the models to be simulated.

## 3. Physical mechanisms and effective parameters

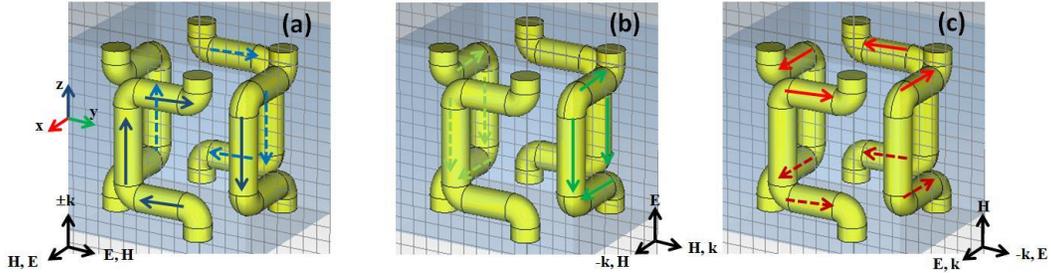

Fig. 2. Schematics of three different magnetic dipoles excited inside the unit cell of the metamaterial under different incidences of light. Arrows indicate magnetic dipole loop currents induced by (a) normally incident light, laterally incident (b) TE-polarized light and (c) TM-polarized light. In all the cases the metamaterial is periodic in the planes perpendicular to and finite along the incident wave vector **k**. **E** and **H** stands for incident electric and magnetic field, respectively. For clarity, only the loop currents excited by **H** perpendicular to the plane of the loops are shown [e.g., the loop current in (a) is excited by **H** along the x-direction]. Using symmetry considerations, current loops at different planes can be also excited [e.g., in (a) if **H** is chosen along the y-direction, magnetic dipoles in the x-z planes are excited, thus the structure becomes polarization independent in the x-y plane for normal incidence.

Different magnetic dipoles [70] with close resonance frequencies can be excited inside the structure in Fig. 1 under different incidences of light. The resonances underlying the magnetic dipoles arise from the combined capacitance and inductance of the nanocircuits inside the structure similar to split-ring-resonators. Fig. 2 schematically illustrates three different magnetic dipoles that can be excited under different incidences of light. Color arrows represent the current loops for the induced magnetic dipoles. The planes of these current loops are perpendicular to the respective incident magnetic field **H**, since the dipoles are magnetically excited. Below we refer to the current loops in the y-z plane (see Fig. 2a) and the x-z plane (see Fig. 2b) as column loops, and the current loops in the x-y plane (see Fig. 2c) as the joist loops. In particular, Fig. 2a illustrates the magnetic dipoles excited by normal incidence. Here, we define the normal incidence such that the incident wave vector **k** is along the z-direction and the structure is periodic in the x-y plane. This is the simplest configuration for fabrication with DLW, where the structures are grown on the substrate parallel to the x-y plane, and subsequent optical characterization. Optical response under this configuration is polarization-independent in the x-y plane. On the other hand, Fig. 2b illustrates the magnetic dipoles excited by TE-polarized laterally incident light. We define the lateral incidence such that the incident **k**-vector lies in the x-y plane parallel to the substrate and the structure is periodic in the plane perpendicular to the **k**-vector. In this case, the TE-polarized light is described as the electromagnetic field with fixed electric field **E** along the z-direction. In contrast, for the TM-polarized laterally incident light electric field **E** is replaced with **H** as shown in Fig. 2c where the corresponding magnetic dipoles are also illustrated.

### A. Effective parameters and field distribution for different incidences of light

In this part, we verify the induced magnetic dipoles illustrated schematically in Fig. 2 based on calculated current density distributions and show the results for the retrieved effective optical parameters for single layer metamaterial structures.

First, we consider the configuration in Fig. 2a where the structure interacts with normally incident light. Fig. 3a shows the resultant transmittance (T), reflectance (R), and absorbance (A). In Fig. 3b, we plot the retrieved effective refractive index, $n = n' + in''$. Retrieved effective permittivity, $\varepsilon_y = \varepsilon_y' + i\varepsilon_y''$, and permeability, $\mu_x = \mu_x' + i\mu_x''$, are shown in Figs. 3c and d, respectively. Notice that a magnetic resonance with a Lorentzian-like lineshape [71, 72] (see Fig. 3d) appears around 26THz and $\mu_x'$ is negative between 26THz and 30THz. The ratio of the vacuum wavelength to unit cell size in the propagation direction (i.e., λ/a ratio) is about 5, which is reasonably large for homogenous effective medium approximation, near the magnetic resonance. Figs. 3e and f show the current density distribution at f=27THz (i.e., near the magnetic resonance frequency). This verifies the column loops illustrated in Fig. 2a.

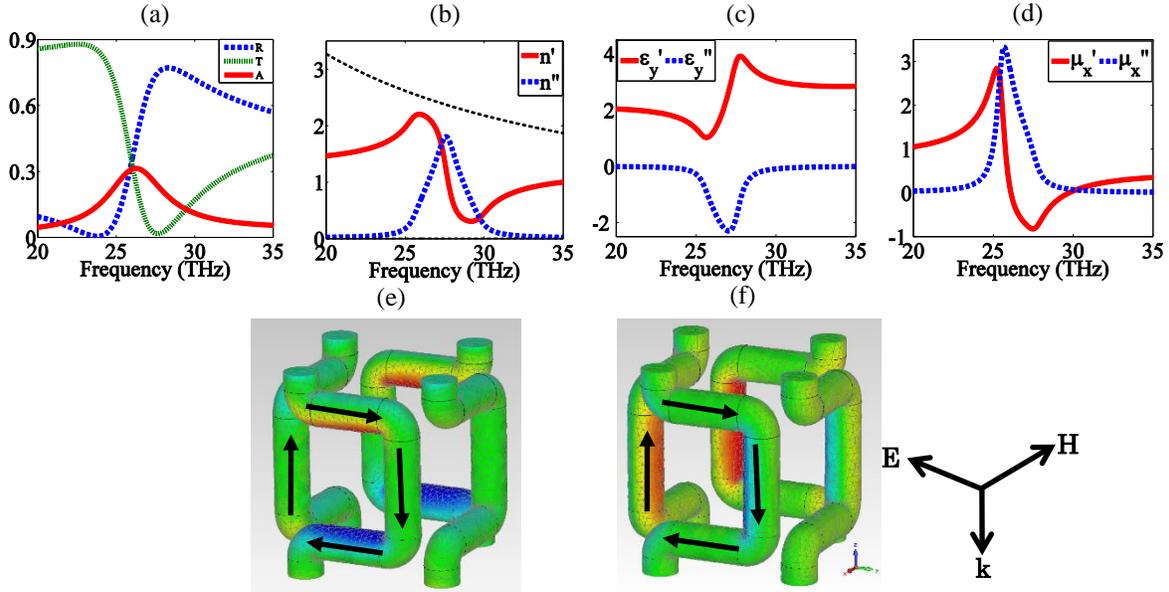

Fig. 3. (**a**) Transmittance (T), reflectance (R), and absorbance (A) spectra for normally incident light. The structure is periodic in the x-y plane and has one layer along the z-direction. Retrieved effective (**b**) refractive index, (**c**) permittivity, and (**d**) permeability. The dashed black line in (b) is the first Brillouin zone edge. (**e**) Current density distribution at f=27THz. Colors show the y-component of the current density **J_y** (i.e., red and blue correspond to the +y- and −y-directions, respectively). (**f**) The same as (e) except that colors show the z-component of the current density **J_z** (i.e., red and blue correspond to the +z- and −z-directions, respectively). Arrows overlaying the surface plots in (e) and (f) indicate the complete dipole loop current. Only one of the loops is shown for clarity (see Fig. 2a for the location of other loop).

When the structure interacts with laterally incident light such as in Figs. 2b and c, then two other magnetic resonances originate depending on the polarization of incident light. Fig. 4 shows the case for the TE-polarized light. There exists magnetic resonance around 22THz. In this case, λ/a ratio is also about 5. The resultant current density distribution, which verifies the column loops illustrated in Fig. 2b, is shown in Figs. 4e and f. Finally, Fig. 5 shows the results for the TM-polarized light. Particularly, Figs. 5e and f show the current density distribution at 40THz near the magnetic resonance, which verifies the joist loops illustrated in Fig. 2c. For this case λ/a ratio is about 3. Although the structure might not seem to be sufficiently subwavelength under this configuration, we should note that the results are still reliable, because (i) no discontinuities are observed in the retrieved results and (ii) the retrieved refractive index is below the first Brillouin zone edge.

### B. Hyperbolic dispersion

The magnetic dipoles discussed above can be used to obtain hyperbolic dispersion. As an example we choose the magnetic dipoles in Fig. 3. We start with considering the TE-polarized electromagnetic waves propagating in the x-z plane. The electric field is fixed along the y-direction. When the incident **k**-vector changes its direction from the z-direction to the x-direction, the corresponding incident field configuration changes from Fig. 3 to Fig. 5. Therefore, one might expect hyperbolic dispersion around the region where $\mu_x' < 0$ (see Fig. 3d) since $\mu_z' > 0$ in the same region. However, we show below that imaginary parts also have important contribution to the dispersion of the metamaterial.

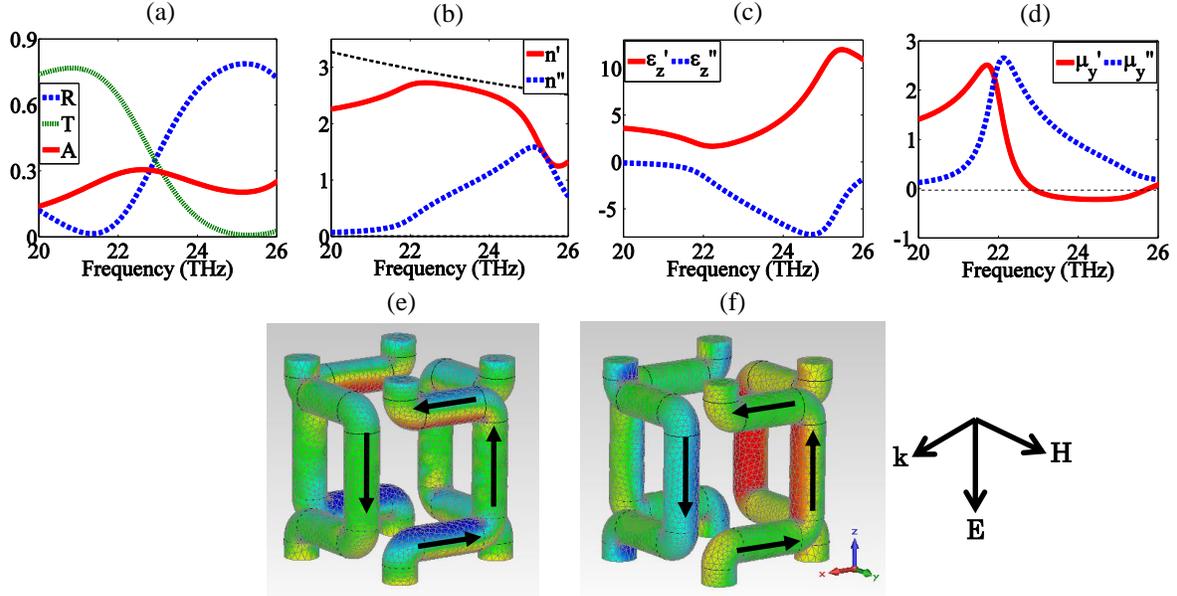

Fig. 4. (a) Transmittance (T), reflectance (R), and absorbance (A) spectra for the TE-polarized laterally incident light. The structure is periodic in the y-z plane and has one layer along the x-direction. Retrieved effective (b) refractive index, (c) permittivity, and (d) permeability. The dashed black line in (b) is the first Brillouin zone edge. (e) Current density distribution at f=22THz. Colors show the x-component of the current density $J_x$ (i.e., red and blue correspond to the +x- and −x-directions, respectively). (f) The same as (e) except for $J_z$. Arrows overlaying the surface plots in (e) and (f) indicate the complete dipole loop current. Only one of the loops is shown for clarity (see Fig. 2b for the location of other loop).

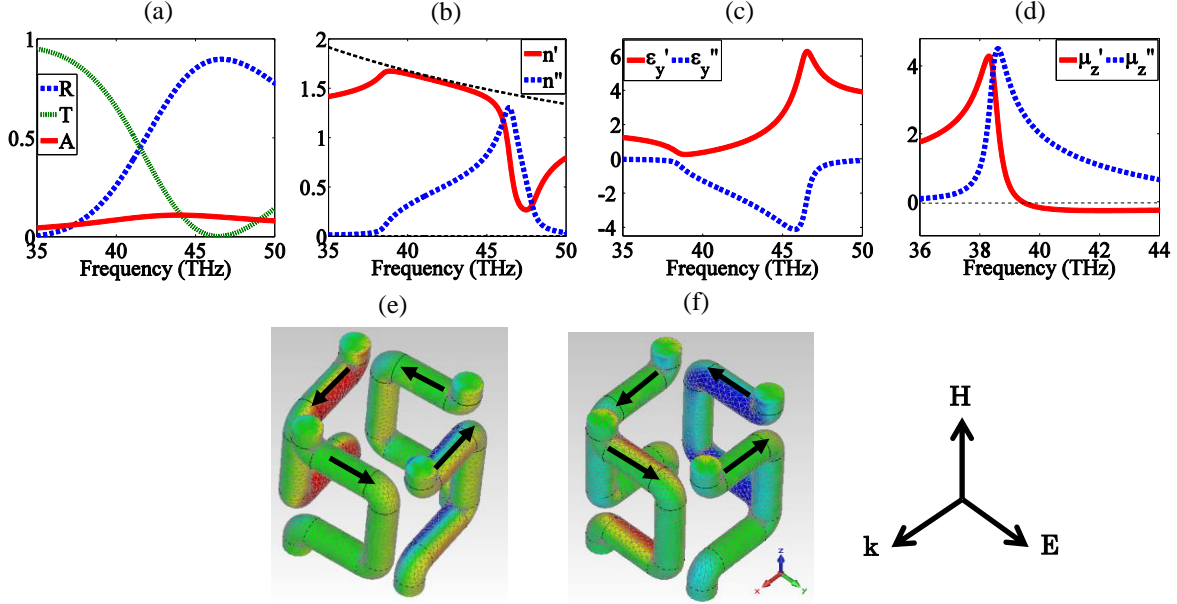

Fig. 5. (a) Transmittance (T), reflectance (R), and absorbance (A) spectra for the TM-polarized laterally incident light. The structure is periodic in the y-z plane and has one-layer along the x-direction. Retrieved effective (b) refractive index, (c) permittivity, and (d) permeability. The dashed black line in (b) is the first Brillouin zone edge. (e) Current density distribution at f=40THz for (e) $J_x$ and (f) $J_y$. Arrows overlaying the surface plots in (e) and (f) indicate the complete dipole loop current. Only one of the loops is shown for clarity (see Fig. 2c for the location of other loop).

In order to demonstrate the hyperbolic dispersion relation, we calculate the tangential and normal components of the effective wave vector inside the metamaterial which are defined as $k_x$ and $k_z$, respectively. The inverted Fresnel formula [69, 73] is used to obtain $k_z$:

$$k_z d = \pm cos^{-1}\left(\frac{1-r^2+t^2}{2t}\right) + 2\pi m \qquad (1)$$

where $d$ is the unit cell thickness along the propagation direction. $r$ and $t$ are the reflection and transmission coefficients, respectively. $m$ is the branch number. The sign is chosen to guarantee a positive imaginary part of $k_z$ and $m$ is selected to promise continuous real part of $k_z$. On the other hand, at the vacuum-metamaterial interface the tangential components of the wave vectors are continuous. Thus, $k_x$ can be expressed as $k_x =$

$k_0 sin\theta$, where $k_0$ is the wave number in free space and $\theta$ is the angle of incidence with respect to the surface normal. Different angles of incidence are set up in the CST simulations and corresponding reflection and transmission coefficients along with $k_x$ values are obtained. $k_z$ values are then calculated from Eq. (1). Finally, based on different pairs of $k_z$ and $k_x$, we obtain the equifrequency contours describing the dispersion for the metamaterial.

The equifrequency contours for one layer structure (i.e., single unit cell along the z-direction and infinite in the x-y plane) gives elliptical dispersion contrary to anticipated hyperbolic dispersion due to the contribution of relatively large imaginary parts of effective optical parameters. However, increasing the number of layers of the proposed metamaterial structure leads to a transition from elliptical dispersion to hyperbolic dispersion. For example, the equifrequency contours in Fig. 6 corresponds to the five-layer structure (i.e., five unit cells along the z-direction and infinite in the x-y plane) which manifests hyperbolic dispersion. The blue, green, red, and cyan lines represent the frequencies of 15THz, 16THz, 18THz, and 20 THz, respectively.

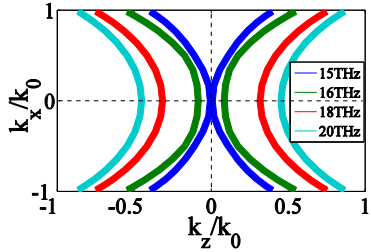

Fig. 6. Equifrequency contours for the TE-polarized light propagating in the x-z plane show hyperbolic dispersion. The structure is periodic in the x-y plane and has five layers along the z-direction

In Figs. 7 and 8, we show the effective parameters for two five-layer structures with different configurations. Fig. 7 shows effective permittivity and permeability under normal incidence with the same geometric configuration as in Fig. 6. Comparing Fig. 7 with Fig. 3, we notice that the magnetic resonance becomes weaker and red shifts from about 26THz to 20THz with increasing number of layers. The structure homogenizes rather slowly as can be seen from relatively different effective parameters compared to single layer.

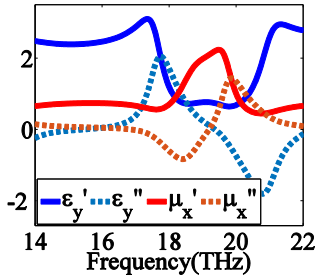

Fig. 7. Effective permittivity and permeability for the metamaterial under normal incidence with five layers along the z-direction and periodic in the x-y plane.

Fig. 8 shows effective permittivity and permeability for the five-layer structure under y-polarized laterally incident light with the same geometric configuration as in Fig. 5 except that there exists five layers along the x-direction. Around 20THz, we observe an electric resonance, which does not appear in one layer structure. This suggests that unlike the magnetic resonances, the electric resonance arises from the interaction between neighboring unit cells rather than intra-unit cell effect.

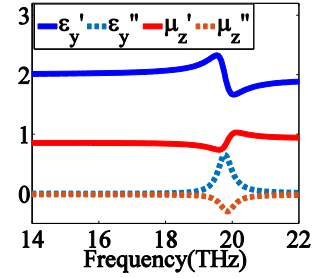

Fig. 8. Effective permittivity and permeability for the metamaterial under lateral incidence with five layers along the x-direction and periodic in the y-z plane.

In the following we show that the retrieved effective parameters in Figs. 7 and 8 are consistent with the hyperbolic dispersion displayed in Fig. 6. For the TE-polarization considered in Fig. 6, because **E** is along the y-direction and the vectors **k** and **H** are in the x-z plane, only $\varepsilon_y$, $\mu_x$, and $\mu_z$ contribute to hyperbolic dispersion, which can be expressed as:

$$\frac{k_x^2}{\varepsilon_y \mu_z} + \frac{k_z^2}{\varepsilon_y \mu_x} = \frac{\omega^2}{c^2} \qquad (2)$$

We consider, for example, f=20THz in Fig. 6. From retrieved results in Figs. 7 and 8, $\varepsilon_y = 0.6869 - i0.7442$, $\mu_x = 1.006 + i1.345$, $\mu_z = 1.016 - i0.2175$. The negative imaginary parts in $\varepsilon_y$ and $\mu_z$ are due to inherent periodicity effects commonly seen in metamaterials [74]. First, considering $\theta = 0°$ (i.e., normal incidence), in Eq. (2) we substitute $k_x = 0$, $k_z$ calculated from Eq. (1), and the retrieved parameters, $\varepsilon_y$ and $\mu_x$ (see Fig. 7), we find the corresponding complex frequency as $19.912 + i1.3258$ THz, which is very close to 20THz.

When $\theta \neq 0°$, $\mu_z$ is also needed. However, the retrieved $\mu_z$ in Fig. 8 does not accurately describe the equifrequency contours in Fig. 6, because the structure corresponding to Fig. 6 has infinite unit cells in the x-y plane and only 5 unit cells along the z-axis. In contrast, the structure corresponding to Fig. 8 has 5 unit cells along the x-axis and infinite unit cells in the y-z plane. Although the numbers of unit cells in different directions are different, the optical properties of these two structures should be qualitatively similar since they are two different pieces of the identical bulk metamaterial. Therefore, starting with the above $\mu_z$ as a guess value and feeding this into Eq. (2) we find through an iterative process that the best fitting value is $\mu_z = 1.75 - i0.6175$. Although the real part is relatively large compared to $\mu_z$ in Fig. 8, the negative imaginary part implies that electric resonance seems to persist.

## 4. Discussion

Based on above analysis, the imaginary part of the effective parameters plays an important role on the type of metamaterial dispersion. In the previous reports (see, for example [51]), since the imaginary parts are usually small compared to the real parts, considering only the real parts of the effective parameters alone are usually sufficient to predict the type of metamaterial dispersion. While, here, the imaginary parts of the effective parameters are in the same order as the real parts or even larger. Therefore, contrary to the metamaterials with low imaginary parts in the effective parameters, one-layer structure, here, results in elliptical dispersion with $\varepsilon'_y > 0$, $\mu'_x < 0$, and $\mu'_z > 0$ and five-layer structure results in hyperbolic dispersion with $\varepsilon'_y > 0$, $\mu'_x > 0$, and $\mu'_z > 0$. We should note that, in the latter, especially negative imaginary part of $\mu_z$ around electric resonance as a result of periodicity effects has important contribution to the transition from elliptical to hyperbolic dispersion. Despite this sharp transition in optical properties as the monolayer structure is transformed into multiple-layer stack (i.e., this can be regarded analogous to graphene versus graphite), the structure slowly homogenizes with increasing number of layers and approaches a bulk hyperbolic metamaterial.

The metamaterial structures, here, were designed specifically for fabrication with DLW processes and subsequent metallization. The functional optical metamaterials resulting from this fabrication approach are usually free-standing structures in air [57, 58, 75]. Therefore, vacuum was selected as a background material in our simulations. Direct metallization in dielectric host media using DLW is also possible [76, 77]. However, no functional optical metamaterial with this approach has been shown.

If the meandering wires are embedded in dielectric media with larger refractive index than vacuum, we find that the resonances and effective material properties red-shift. Therefore, it is expected that the hyperbolic dispersion should also red-shift with larger refractive index. Considering the underlying resonant magnetic dipole modes (see Fig. 2) the red-shift in magnetic resonances in Figs. 3-5, for example, can be easily explained by a simple LC circuit model [18, 71, 78]. Effectively, the meandering wires in Figs. 3 and 4 behave similar to two-gap split-ring-resonators (SRRs) and the meandering wires in Fig. 5 behave similar to a four-gap SRR. Embedding these SRR-like structures inside host media with larger refractive index than vacuum results in increase in equivalent circuit capacitance, hence red-shift in resonance frequency.

On the other hand, if we decrease the length or the diameter of the wires, blue-shift occurs in optical magnetic response, because decreasing the length of the wires reduce the equivalent circuit impedance and decreasing the diameter of the wires by keeping the wire positions fixed reduce the equivalent circuit capacitance due to larger gaps. Thus, in both cases magnetic resonance frequency blue-shifts consistent with LC circuit model.

We should note that the retrieved effective parameters above are obtained by inverting transmission and reflection coefficients in accordance with homogeneous effective medium (HEM) approximation discussed in [69, 73, 74]. This retrieval procedure uniquely returns the impedance ($z = z' + iz''$) and $n''$ by making use of the physical requirements that $z' > 0$ and $n'' > 0$ for passive material. However, there exists an ambiguity in determining $n'$ due to multiple solutions. Once this ambiguity is resolved, the effective permittivity ($\varepsilon$) and permeability ($\mu$) are determined from $\varepsilon = n/z$ and $\mu = nz$, respectively, without any constraint on imaginary parts of $\varepsilon$ and $\mu$. In our case, $n'$ was obtained from continuous 0[th] order branch under the first Brillouin zone edge, which in turn was verified by (1) the resultant resonances in effective constitutive parameters that are consistent with the field distributions and (2) multiple-layer simulations.

The origin of resultant commonly observed negative imaginary parts in the retrieved constitutive parameters under the HEM approximation has been extensively interrogated [74, 79-84]. It was shown that these negative imaginary parts near the resonances arise from inherent periodicity of the metamaterial if the actual periodic metamaterial structure is approximated by a HEM with the same scattering parameters as the periodic structure [74, 85]. However, every inhomogeneous medium exhibits spatial dispersion (i.e., polarization and magnetization at a given location depends on spatial distribution of the fields) [82], which is not considered in the HEM approximation. Further studies have shown that the incorporation of spatial dispersion improves the accuracy of the effective parameter retrieval procedure by removing the periodicity artifacts such as negative imaginary parts in constitutive parameters [82-84].

Finally, in Fig. 9 we plot in the same graph the effective permeability values corresponding to three different magnetic dipoles illustrated in Fig. 2 and studied in Figs. 3-5. It is worth mentioning that three effective permeability values intersect non-trivially around 40THz (i.e., convergence below 20THz is uninteresting due to asymptotic non-magnetic response at low frequencies). This shows that the structure has potential for three-dimensionally isotropic permeability despite geometric anisotropy.

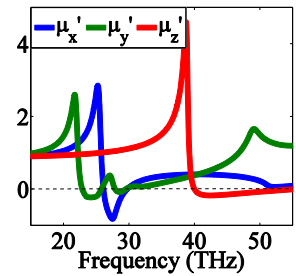

Fig. 9. Real parts of the effective magnetic permeability in the same graph corresponding to three different magnetic dipoles studied in Figs. 3-5. The curves intersecting around 40THz show potential for non-trivial three-dimensionally isotropic permeability.

Despite operating at a single frequency, considering the difficulty of making isotropic metamaterials at optical frequencies, this result is still interesting. Moreover, the structure may be optimized to operate over a wider bandgap. Tunability with the incorporation of for example, liquid crystals [86-90], can also be utilized to mitigate the bandwidth issue.

## 5. Conclusion

In summary, we propose a hyperbolic metamaterial structure operating at mid-infrared frequencies. The structure is feasible to fabricate using combined state-of-the-art STED-DLW technologies and subsequent electroplating of gold. The structure supports three different magnetic dipoles. These dipoles cannot only contribute to hyperbolic dispersion but also can provide opportunity to design three-dimensionally isotropic effective permeability. Additionally, we discuss the influence of the imaginary parts of the effective parameters on metamaterial dispersion which is usually overlooked in the literature.

## Acknowledgments

The work at Michigan Technological University was supported by the National Science Foundation under grant ECCS-1202443. We would like to thank Costas M. Soukoulis for fruitful discussions on meandering wire structures and Martin Wegener for discussions on fabrication of the structures with STED-DLW processes.

References

1. V. G. Veselago, "The electrodynamics of substances with simultaneously negative values of ε and μ," Sov. Usp. **10**, 509-514 (1968).
2. J. B. Pendry, "Negative refraction makes a perfect lens," Phys. Rev. Lett. **85**, 3966 (2000).
3. Z. Jacob, L. V. Alekseyev, and E. Narimanov, "Optical hyperlens: far-field imaging beyond the diffraction limit," Opt. Express **14**, 8247-8256 (2006).
4. Z. Liu, H. Lee, Y. Xiong, C. Sun, and X. Zhang, "Far-field optical hyperlens magnifying sub-diffraction-limited objects," Science **315**, 1686 (2007).
5. X. Zhang and Z. Liu, "Superlenses to overcome the diffraction limit," Nature Materials **7**, 435-441 (2008).
6. J. Rho, Z. Ye, Y. Xiong, X. Yin, Z. Liu, H. Choi, G. Bartal, and X. Zhang, "Spherical hyperlens for two-dimensional sub-diffractional imaging at visible frequencies," Nature Commun. **1**, 143 (2010).
7. J. B. Pendry, D. Schurig, and D. R. Smith, "Controlling electromagnetic fields," Science **312**, 1780-1782 (2006).
8. U. Leonhardt, "Optical conformal mapping," Science **312**, 1777-1780 (2006).
9. D. Schurig, J. J. Mock, B. J. Justice, S. A. Cummer, J. B. Pendry, A. F. Starr, and D. R. Smith, "Metamaterial electromagnetic cloak at microwave frequencies," Science **314**, 977-980 (2006).
10. N. I. Landy, S. Sajuyigbe, J. J. Mock, D. R. Smith, and W. J. Padilla, "Perfect metamaterial absorber," Phys. Rev. Lett. **100**, 207402 (2008).
11. K. Aydin, V. E. Ferry, R. M. Briggs, and H. A. Atwater, "Broadband polarization-independent resonant light absorption using ultrathin plasmonic super absorbers," Nature Commun. **2**, 517 (2011).
12. U. Leonhardt and T. G. Philbin, "Quantum levitation by left-handed metamaterials," New J. Phys. **9**, 254 (2007).
13. D.O. Guney and D.A. Meyer, "Negative refraction gives rise to the Klein paradox," Phys. Rev. A **79**, 063834 (2009).
14. D.A. Genov, S. Zhang, and X. Zhang, "Mimicking celestial mechanics in metamaterials," Nature Phys. **5**, 687-692 (2009).
15. I. I. Smolyaninov and E. E. Narimanov, "Metric signature transitions in optical metamaterials," Phys. Rev. Lett. **105**, 067402 (2010).
16. I. Bulu, H. Caglayan, K. Aydin, and E. Ozbay, "Compact size highly directive antennas based on the SRR metamaterial medium," New J. Phys. **7**, 223 (2005).
17. H. Odabasi, F. Teixeira, and D.O. Guney, "Electrically small, complementary electric-field-coupled resonator antennas," J. Appl. Phys. **113**, 084903 (2013).
18. A. Vora, J. Gwamuri, N. Pala, A. Kulkarni, J. M. Pearce, and D. O. Guney, "Exchanging Ohmic losses in metamaterial absorbers with useful optical absorption for photovoltaics," Sci. Rep., **4**, 4901 (2014).
19. M. I. Aslam and D. O. Guney, "On negative index metamaterial spacers and their unusual optical properties," Progress in Electromagnetics Research B **47**, 203-217 (2013).
20. C. L. Cortes, W. Newman, S. Molesky, and Z. Jacob, "Quantum nanophotonics using hyperbolic metamaterials," J. Opt. **14**, 063001 (2012).
21. A. Poddubny, I. Iorsh, P. Belov, and Y. Kivshar, "Hyberbolic metamaterials," Nature Photon. **7**, 948-957 (2013).
22. P. Shekhar, J. Atkinson, and Z. Jacob, "Hyperbolic metamaterials: fundamentals and applications," arXiv: 1401.2453.
23. R. A. Shelby, D. R. Smith, and S. Schultz, "Experimental verification of a negative index of refraction," Science **292**, 77-79 (2001).
24. M. I. Aslam and D. O. Guney, "Dual band double-negative polarization independent metamaterial for the visible spectrum," J. Opt. Soc. Am. B **29**, 2839-2847 (2012).
25. M. I. Aslam and D. O. Guney, "Surface plasmon driven scalable low-loss negative-index metamaterial in the visible spectrum," Phys. Rev. B **84**, 195465 (2011).
26. M. A. Noginov, H. Li, Y. A. Barnakov, D. Dryden, G. Nataraj, G. Zhu, C. E. Bonner, M. Mayy, Z. Jacob, and E. E. Narimanov, "Controlling spontaneous emission with metamaterials," Opt. Lett. **35**, 1863-1865 (2010).
27. A. N. Poddubny, P. A. Belov, and Y. S. Kivshar, "Spontaneous radiation of a finite-size dipole in hyperbolic media," Phys. Rev. A **84**, 023807 (2011).
28. T. Tumkur, G. Zu, P. Black, Y. A. Barnakov, C. E. Bonner, and M. A. Noginov, "Control of spontaneous emission in a volume of functionalized hyperbolic metamaterial," Appl. Phys. Lett. **99**, 151115 (2011).
29. J. Kim, V. P. Drachev, Z. Jacob, G. V. Naik, A. Boltasseva, E. E. Narimanov, and V. M. Shalaev, "Improving the radiative decay rate for dye molecules with hyperbolic metamaterials," Opt. Express **20**, 8100-8116 (2012).
30. W. D. Newman, C. L. Cortes, and Z. Jacob, "Enhanced and directional single-photon emission in hyperbolic metamaterials," J. Opt. Soc. Am. B **30**, 766-775 (2013).
31. M. Y. Shalaginov, S. Ishii, J. Liu, J. Irudayaraj, A. Lagutchev, A. V. Kildishev, and V. M. Shalaev, "Broadband enhancement of spontaneous emission from nitrogen-vacancy centers in nanodiamonds by hyperbolic metamaterials," arXiv: 1304.6750.
32. E. E. Nerimanov, H. Li, Y. A. Barnakov, T. U. Tumkur, and M. A. Noginov, "Reduced reflection from roughened hyperbolic metamaterial," Opt. Express **21**, 14956-14961 (2013).
33. J. Liu, G. V. Naik, S. Ishii, C. DeVault, A. Boltasseva, V. M. Shalaev, and E. Narimanov, "Optical absorption of hyperbolic metamaterial with stochastic surfaces," Opt. Express **22**, 8893-8901 (2014).
34. L. Ferrari, D. Lu, D. Lepage, and Z. Liu, "Enhanced spontaneous emission inside hyperbolic metamaterials," Opt. Express **22**, 4301-4306 (2014).
35. Y. Guo, C. L. Cortes, S. Molesky, and Z. Jacob, "Broadband super-Planckian thermal emission from hyperbolic metamaterials," Appl. Phys. Lett. **101**, 131106 (2012).
36. Y. He, S. He, and X. Yang, "Optical field enhancement in nanoscale slot waveguides of hyperbolic metamaterials," Opt. Lett. **37**, 2907-2909 (2012).
37. Y. He, S. He, J. Gao, and X. Yang, "Giant transverse optical forces in nanoscale slot waveguides of hyperbolic metamaterials," Opt. Express **20**, 22372-22382 (2012).
38. L. V. Alekseyev, V. A. Podolskiy, and E. E. Narimanov, "Homogeneous hyperbolic systems for terahertz and far-infrared frequencies," Advances Optoelectron. **2012**, 267564 (2012).
39. J. Sun, J. Zhou, B. Li, and F. Kang, "Indefinite permittivity and negative refraction in natural material: graphite," Appl. Phys. Lett. **98**, 101901 (2011).

40. C. M. Soukoulis and M. Wegener, "Optical metamaterials—more bulky and less lossy," Science **330**, 1633-1634 (2010).
41. J. Valentine, S. Zhang, T. Zentgraf, E. Ulin-Avillam D. A. Genov, G. Bartal, and X. Zhang, "Three-dimensional optical metamaterials with negative refractive index," Nature **455**, 376-379 (2008).
42. A. J. Hoffman, L. Alekseyev, S. S. Howard, K. J. Franz, D. Wasserman, V. A. Podolskiy, E. E. Narimanov, D. L. Sivco, and C. Gmachi, "Negative refraction in semiconductor metamaterials," Nature Mat. **6**, 946-950 (2007).
43. X. Yang, J. Yao, J. Rho, X. Yin, and X. Zhang, "Experimental realization of three-dimensional indefinite cavities at the nanoscale with anomalous scaling laws," Nature Photon. **6**, 450-454 (2012).
44. H. N. S. Krishnamoorthy, Z. Jacob, E. Narimanov, I. Kretzschmar, and V. M. Menon, "Topological transitions in metamaterials," Science **336**, 205-209 (2012).
45. T. U. Tumkur, L. Gu, J. K. Kitur, E. E. Narimanov, and M. A. Noginov, "Control of absorption with hyperbolic metamaterials," Appl. Phys. Lett. **100**, 161103 (2012).
46. J. Yao, Z. Liu, Y. Liu, Y. Wang, C. Sun, G. Bartal, A. M. Stacy, and X. Zhang, "Optical negative refraction in bulk metamaterials of nanowires," Science **321**, 930 (2008).
47. M. A. Noginov, Y. A. Barnakov, G. Zu, T. Tunkur, H. Li, and E. E. Narimanov, "Bulk photonic metamaterial with hyperbolic dispersion," Appl. Phys. Lett. **94**, 151105 (2009).
48. J. Kanungo and J. Schilling, "Experimental determination of the principal dielectric functions in silver nanowire metamaterials," Appl. Phys. Lett. **97**, 021903 (2010).
49. L. M. Custodio, C. T. Sousa, J. Ventura, J. M. Teixeira, P. V. S. Marques, and J. P. Araujo, "Birefringence swap at the transition to hyperbolic dispersion in metamaterials," Phys. Rev. B **85**, 165408 (2012).
50. S.M. Prokes, O. J. Glembocki, J. E. Livenere, T. U. Tumkur, J. K. Kitur, G. Zhu, B. Wells, V. A. Podolskiy, and M. A. Noginov, "Hyperbolic and plasmonic properties of Silicon/Ag aligned nanowire arrays," Opt. Express **21**, 14962-14974 (2013).
51. S. S. Kruk, D. A. Powell, A. Minovich, D. N. Neshev, and Y. S. Kivshar, "Spatial dispersion of multilayer fishnet metamaterials," Opt. Express **20**, 15100-15105 (2012).
52. A. Andryieuski, A. V. Lavrinenko, and D. N. Chigrin, "Graphene hyperlens for terahertz radiation," Phys. Rev. B **86**, 121108 (2012).
53. I. V. Iorsh, I. S. Mukhin, I. V. Shadrivov, P. A. Belov, and Y. S. Kivshar, "Hyperbolic metamaterials based on multilayer graphene structures," Phys. Rev. B **87**, 075416 (2013).
54. D. O. Guney, Th. Koschny, M. Kafesaki, and C. M. Soukoulis, "Connected bulk negative index photonic metamaterials," Opt. Lett. **34**, 506-508 (2009).
55. D. O. Guney, Th. Koschny, and C. M. Soukoulis, "Intra-connected three-dimensionally isotropic bulk negative index photonic metamaterial," Opt. Express **18**, 12348-12353 (2010).
56. C.B. Arnold, and A. Pique, "Laser direct-write processing," MRS Bull. **32**, 9-15 (2007).
57. M.S. Rill, C. Plet, M. Thiel, I. Staude, G. von Freymann, S. Linden, and M. Wegener, "Photonic metamaterials by direct laser writing and silver chemical vapour deposition," Nat. Mater. **7**, 543-546 (2008).
58. J.K. Gansel, M. Thiel, M.S. Rill, M. Decker, K. Bade, V. Saile, G. von Freymann, S. Linden, and M. Wegener, "Gold helix photonic metamaterial as broadband circular polarizer," Science **325**, 1513-1515 (2009).
59. M.S. Rill, C.E. Kriegler, M. Thiel, G. von Freymann, S. Linden, and M. Wegener, "Negative-index bianisotropic photonic metamaterial fabricated by direct laser writing and silver shadow evaporation," Opt. Lett. **34**, 19-21 (2009).
60. S. Kawata, H.B. Sun, T. Tanaka, and K. Takada, "Finer features for functional microdevices," Nature **412**, 697-698 (2001).
61. F. Formanek, N. Takeyasu, T. Tanaka, K. Chiyoda, A. Ishikawa, and S. Kawata, "Three-dimensional fabrication of metallic nanostructures over large areas by two-photon polymerization," Opt. Express **14**, 800-809 (2006).
62. S. W. Hell and J. Wichmann, "Breaking the diffraction resolution limit by stimulated emission: stimulated-emission-depletion fluorescence microscopy," Opt. Lett. **19**, 780-782 (1994).
63. T. A. Klar, S. Jakobs, M. Dyba, A. Egner, and S. W. Hell, "Fluorescence microscopy with diffraction resolution barrier broken by stimulated emission," Proc. Natl. Acad. Sci. **97**, 8206-8210 (2000).
64. S. W. Hell, "Strategy for far-field optical imaging and writing without diffraction limit," Phys. Lett. A **326**, 140 (2004).
65. S. W. Hell, "Microscopy and its focal switch," Nat. Methods 6, 24-32 (2009).
66. J. Fischer, T. Ergin, and M. Wegener, "Three-dimensional polarization independent visible frequency carpet invisibility cloak," Opt. Lett. **36**, 2059-2061 (2011).
67. M. Thiel, J. Ott, A. Radke, J. Kaschke, and M. Wegener, "Dip-in depletion optical lithography of three-dimensional chiral polarizers," Opt. Lett. **38**, 4252-4255 (2013).
68. J. Fischer, M. Thiel, and M. Wegener, "Matter made to order," IEEE Spectrum **51**, 34-38 (2014).
69. D. R. Smith, S. Schultz, P. Markos, and C. M. Soukoulis, "Determination of effective permittivity and permeability of metamaterials from reflection and transmission coefficients," Phys. Rev. B **65**, 195104 (2002).
70. D. O. Guney, Th. Koschny, and C. M. Soukoulis, "Surface plasmon driven electric and magnetic resonators for metamaterials," Phys. Rev. B **83**, 045107 (2011).
71. D. O. Guney, Th. Koschny, and C. M. Soukoulis, "Reducing ohmic losses in metamaterials by geometric tailoring," Phys. Rev. B **80**, 125129 (2009).
72. S. Zhang, W. Fan, K. J. Malloy, S. R. J. Brueck, N. C. Panoiu, and R. M. Osgood, "Near-infrared double negative metamaterials," Opt. Express **13**, 4922-4930 (2005).
73. C. Menzel, C. Rockstuhl, T. Paul, F. Lederer, and T. Pertsch, "Retrieving effective parameters for metamaterials at oblique incidence," Phys. Rev. B **77**, 195328 (2008).
74. Th. Koschny, P. Markos, E. N. Economou, D. R. Smith, D. C. Vier, and C. M. Soukoulis, "Impact of inherent periodic structure on effective medium description of left-handed and related metamaterials," Phys. Rev. B **71**, 245105 (2005).
75. X. Xiong, S.-C. Jiang, Y.-H. Hu, R.-W. Pang, and M. Wang, "Structured metal film as perfect absorber," Advanced Materials **25**, 3994-4000 (2013).
76. K. Vora, S. Kang, S. Shukla, and E. Mazur, "Fabrication of disconnected three-dimensional silver nanostructures in a polymer matrix," Appl. Phys. Lett. **100**, 063120 (2012).
77. S. Kang, K. Vora, and E. Mazur, "One-step direct laser metal writing of sub-100nm 3D silver nanostructures in a gelatin matrix," Nanotechnology **26**, 121001 (2015).
78. J. Zhou, Th. Koschny, M. Kafesaki, E. N. Economou, J. B. Pendry, and C. M. Soukoulis, "Saturation of the magnetic

response of split ring-resonators at optical frequencies," Phys. Rev. Lett. **95**, 223902 (2005).
79. L. Zhen, J. T. Jiang, W. Z. Shao, and C. Y. Xu, "Resonance-antiresonance electromagnetic behavior in a disordered dielectric composite," Appl. Phys. Lett. **90**, 142907 (2007).
80. W. Smigaj and B. Gralak, "Validity of the effective-medium approximation of photonic crystals," Phys. Rev. B **77**, 235445 (2008).
81. C. Tserkezis, "Effective parameters for periodic photonic structures of resonant elements," J. Phys: Condens. Matter **21**, 155404 (2009).
82. A. Ludwig and K. J. Webb, "Accuracy of effective medium parameter extraction procedures for optical metamaterials," Phys. Rev. B **81**, 113103 (2010).
83. A. Alu, "Restoring the physical meaning of metamaterial constitutive parameters," arXiv: 1012.1353.
84. A. Alu, "First-principles homogenization theory for periodic metamaterials," Phys. Rev. B **84**, 075153 (2011).
85. T. Koschny, P. Markos, D. R. Smith, and C. M. Soukoulis, "Resonant and antiresonant frequency dependence of the effective parameters of metamaterials," Phys. Rev. E **68**, 065602(R) (2003).
86. Q. Zhao, L. Kang, B. Du, B. Li, J. Zhou, H. Tang, X. Liang, and B. Zhang, "Electrically tunable negative permeability metamaterials based on nematic liquid crystals," Appl. Phys. Lett. **90**, 011112 (2007).
87. D. H. Werner, D.-H. Kwon, I.-C. Khoo, A. V. Kildishev, and V. M. Shalaev, "Liquid crystal clad near-infrared metamaterials with tunable negative-zero-positive refractive indices," Opt. Express **15**, 3342-3347 (2007).
88. M. V. Gorkunov and M. A. Osipov, "Tunability of wire-grid metamaterial immersed into nematic liquid crystal," J. Appl. Phys. **103**, 036101 (2008).
89. I. C. Khoo, "Nonlinear optics, active plasmonic and tunable metamaterials with liquid crystals," Progress in Quantum Electronics **38**, 77–117 (2014).
90. J. Ptasinski, I.-C. Khoo, and Y. Fainman, "Enhanced optical tuning of modified-geometry resonators clad in blue phase liquid crystals," Opt. Lett. **39**, 5435-5438 (2014).